\documentclass[conference]{IEEEtran}
\IEEEoverridecommandlockouts
\usepackage{cite}
\usepackage{amsmath,amssymb,amsfonts}
\usepackage{algorithmic}
\usepackage{graphicx}
\usepackage{textcomp}
\usepackage{xcolor}
\usepackage{svg}
\usepackage{orcidlink}
\usepackage{booktabs} 
\usepackage{xcolor,colortbl} 
\usepackage{subcaption}
\usepackage{float}
\usepackage[margin=1in]{geometry}
\usepackage{stfloats}
\usepackage{xcolor}
\usepackage{longtable}
\usepackage{array}
\usepackage{multirow}
\usepackage{array,booktabs}
\def\BibTeX{{\rm B\kern-.05em{\sc i\kern-.025em b}\kern-.08em
    T\kern-.1667em\lower.7ex\hbox{E}\kern-.125emX}}

\usepackage{fancyhdr}
\usepackage{geometry}
\usepackage{lipsum} 

\begin{document}

\title{Green Wave as an Integral Part for the Optimization of Traffic Efficiency and Safety: A Survey}

\author{
\IEEEauthorblockN{
Kranthi Kumar Talluri\IEEEauthorrefmark{1}\IEEEauthorrefmark{4}, Christopher Stang\IEEEauthorrefmark{2}\IEEEauthorrefmark{4}, Galia Weidl\IEEEauthorrefmark{1} 
}

\IEEEauthorblockA{\IEEEauthorrefmark{1}Aschaffenburg University of Applied Sciences, Aschaffenburg, 63743 Germany}
\IEEEauthorblockA{\IEEEauthorrefmark{2}ZF Friedrichshafen AG, Safety Mobility Simulation,
        Friedrichshafen, Germany}
\IEEEauthorblockA{\IEEEauthorrefmark{4}These authors contributed equally to this work.}
\IEEEauthorblockA{Email: 
\IEEEauthorrefmark{1}\{KranthiKumar.Talluri, Galia.Weidl\}@th-ab.de,
\IEEEauthorrefmark{2}christopher.stang@zf.com}

}

\maketitle

\begin{abstract}
Green Wave provides practical and advanced solutions to improve traffic efficiency and safety through network coordination. Nevertheless, the complete potential of Green Wave systems has yet to be explored. Utilizing emerging technologies and advanced algorithms, such as AI or V2X, would aid in achieving more robust traffic management strategies, especially when integrated with Green Wave. This work comprehensively surveys existing traffic control strategies that enable Green Waves and analyzes their impact on future traffic management systems and urban infrastructure. Understanding previous research on traffic management and its effect on traffic efficiency and safety helps explore the integration of Green Wave solutions with smart city initiatives for effective traffic signal coordination. This paper also discusses the advantages of using Green Wave strategies for emission reduction and considers road safety issues for vulnerable road users, such as pedestrians and cyclists. Finally, the existing challenges and research gaps in building robust and successful Green Wave systems are discussed to articulate explicitly the future requirement of sustainable urban transport.
\end{abstract}


\section{Introduction} 
\label{sec:intro}
Urban cities face a significant economic loss due to the growing challenges of traffic congestion. An estimated billions of dollars are spent annually by cities as a consequence of fuel wastage, production loss, infrastructure damage, and vehicle maintenance costs \cite{li2024green,shi2020coordination}. As the urban population and vehicle ownership are expanding, the seriousness of this issue is growing drastically, creating an unusual burden on current transportation systems. Traditional traffic management networks fail to cope with increasing traffic pattern complexities, mainly when critical congestion levels are reached during peak travel hours \cite{zheng2020novel, talluri2024bayesian, talluri2024impact}.

The need for evolved traffic management strategies is increasing rapidly due to the advancements in building smart cities through the Internet of Things (IoT) \cite{chen2016cooperative}. Innovative and modern solutions enhance traffic flow, improve road safety, and mitigate traffic congestion by leveraging real-time traffic data patterns and incorporating automated decision-making capabilities using AI/ML techniques \cite{zhang2021solving}. It aims to achieve a sustainable environment, urban advancement, and immediate response during emergencies through proactive, intelligent decision-making and adaptive transportation systems. These systems are capable of adjusting traffic signal timings in real-time based on changing traffic conditions, which was not possible with traditional traffic management solutions. Cutting-edge approaches \cite{wang2021collaborative}, particularly Green Wave, create synchronized infrastructures that enable vehicles to travel across multiple intersections. Green Wave achieves uninterrupted traffic flow by maintaining optimal speed without redundant stops and starts \cite{yang2020study, chochliouros2021c}. Its fundamental principle is to facilitate the smooth flow of platoons of vehicles through coordinated traffic signal corridors \cite{gao2023proactive}. When integrated with emerging smart city technologies, Green Wave implementations become more robust due to the possibility of building advanced algorithms and processing real-time data to make intelligent, dynamic decisions about traffic flow management \cite{fickas2021green}.

Environmental considerations are becoming fundamental to Green Wave implementation for the optimized flow of traffic, as the latest research works highlight their ability to achieve low fuel consumption and reduced vehicle emissions \cite{zhao2021modeling}, \cite{kiers2017effect}.  The traditional stop-and-go ideology of urban traffic systems reduces efficient engine operation due to repeated acceleration after stopping. Green Wave reduces these effects through smooth vehicle movement at optimal speed. Based on research \cite{de2011traffic, fazzini2022effects}, Green Wave and smart intersections also provided a significant enhancement in air quality with a reduction in carbon and other pollutant emissions at coordinated corridors.

Rising technologies, such as connected vehicles and smart infrastructure, enhance Green Wave system implementations. Precise coordination and optimal traffic flow can be achieved through enabling communication between vehicles and infrastructures using vehicle-to-infrastructure (V2I), vehicle-to-roadside (V2R), and vehicle-to-vehicle (V2V) technologies \cite{kui2019eco, islam2018impact}. Moreover, drivers can be guided to maintain a real-time optimal speed, ensuring smooth driving through green lights at every intersection. Bidirectional communication prospects of V2X in transmission and reception of information about location, speed, and destination by vehicles help achieve sophisticated traffic management through Green Wave \cite{gao2023proactive, yang2020study, chochliouros2021c}. The resulting connected vehicle infrastructure offers a dynamic solution that can effectively respond to modern traffic conditions.


Furthermore, safety is emerging as a critical aspect of the Green Wave in urban environments featuring diverse road users \cite{yuan2022deep}. Intricate safety protocols are incorporated into advanced systems to protect pedestrians and cyclists, ensuring the smooth flow of emergency vehicles without affecting vehicle traffic flow \cite{cao2022gain, de2019green}. Balancing the desire for smooth traffic flow, safe pedestrian crossing times, and prioritizing emergency vehicle flow is crucial for these systems. Considering all the aforementioned points, the contributions of this work are as follows:
\begin{itemize} 
    \item Conducted a comprehensive review on traffic signal control and Green Wave systems and their impact on traffic efficiency, environment, and safety, as limited surveys exist in this area.
    \item Explored the advantage of integrating the latest technologies like V2X, 5G, and edge computing to achieve enhanced Green Wave systems. 
    \item Identified future trends and challenges in real-time scalability, data privacy, and integration of vulnerable road users.    
\end{itemize}

The rest of the paper is organized as follows. Section II discusses traffic signal control with a focus on Green Wave implementation, and Section III briefly describes current trends and the impact of Green Waves on the environment and safety. This is followed by challenges and future directions for implementing the Green Wave technology in section IV. The subsequent reviews of the different sections are based on a selection of scientific articles identified according to the features listed in Table \ref{tab:combined_literature}. In the search process, keywords were combined using the Boolean OR operator. Further literature necessary for explanations may go beyond the search terms.

\begin{table}[h]
\renewcommand{\arraystretch}{1.2}
\centering
\caption{Key Search parameters of literature review}
\begin{tabular}{|>{\centering\arraybackslash}m{0.9cm}|
                 >{\centering\arraybackslash}m{1.5cm}|
                 >{\centering\arraybackslash}m{1.8cm}|
                 >{\centering\arraybackslash}m{2.5cm}|}
\hline
\textbf{Section} & \textbf{Date Range} & \textbf{Database} & \textbf{Keywords}\\
\hline
\multirow{5}{*}{II} & \multirow{5}{*}{2018-2024} 
                   & \multirow{5}{*}{\shortstack{ \\IEEE Xplore \\ Scopus \\ Google Scholar}} 
                   & Green Wave \\ 
                   &&& Traffic light control \\
                   &&& Traffic signal control \\
                   &&& Traffic signal optimization \\ 
                   &&& Intelligent traffic light \\
\hline
\multirow{5}{*}{III} & \multirow{5}{*}{2011-2024} 
                    & \multirow{5}{*}{\shortstack{ \\IEEE Xplore \\ Scopus \\ Google Scholar}} 
                    & Green Wave \\ 
                    &&& V2X in Green Wave \\
                    &&& Green Wave impact on environment \\
                    &&& Green Wave impact on safety \\ 
                    &&& Strategies for Green Wave \\
\hline
\end{tabular}
\label{tab:combined_literature}
\end{table}

\section{Methods of Traffic Signal Control For Multiple Intersection Coordination and Green Wave Implementation} 
\label{sec:Methods of Traffic Signal Control}
The growing urbanization and resulting increase in traffic volume necessitate intelligent traffic signal control strategies to enhance traffic safety and efficiency. To meet both objectives, various methods have been developed in recent years. From a methodological point of view, the control strategies are split into \textit{Fixed-Time}, \textit{Actuated}, and \textit{Adaptive and Intelligent Signal Control} \cite{wang2021traffic, warberg2008green, miletic2022review, miao2021survey, eom2020traffic, wang2022critical, ranpura2022review}. Based on the keywords identified in the literature review for this section, the analysis will highlight the evolution and emerging trends in control strategies, with a focus on multiple-intersection coordination and the feasibility of Green Waves. As the authors give a high-level overview of the different methods and their key technical developments, further explanations can be found in the corresponding references. For the sake of clarity, commonly used terminology in the field of traffic control will be explained in advance.
\begin{itemize}
    \item Phase: represents a specific combination of traffic signal states, ensuring conflict-free movements in any direction.
    \item Cycle: The chronological and repeating order of all phases represents a cycle. The duration of one complete cycle is called cycle time.
    \item Green split: corresponds to the amount of green time per cycle.
\end{itemize}



\subsection{Fixed-Time Signal Control}
\label{sec:Fixed-Time Signal Control}
Traditional traffic signal control uses predefined schedules for the sequence and duration of traffic signal phases. Those schedules are built on historical data and under the assumption of a stable traffic demand at different times of the day \cite{wang2021traffic, warberg2008green, miletic2022review, miao2021survey, eom2020traffic, yue2021evolution, wang2022critical, ranpura2022review}. As static methods rely on offline optimization, they are cost-effective and easy to implement in real-world applications. In this respect, the method of \textit{Webster} \cite{webster1958traffic} was a milestone in traffic control design, representing a mathematical model for minimizing the delay at a single intersection. This is realized by calculating the optimal cycle time and green split in consideration of traffic flow data and intersection saturation \cite{webster1958traffic, wei2019survey}. To maximize traffic efficiency, coordination is necessary along multiple intersections, resulting in a Green Wave along a specific infrastructure corridor. The implementation of Green Waves reduces the number of stops for the affected traffic participants by calculating an offset \(\Delta t_{i,j}\) between the fixed cycle phases of two adjacent intersections \(i\) and \(j\) with a distance of \(L_{i,j}\) \cite{miletic2022review, wei2019survey}.
\begin{equation}
    \Delta t_{i,j}=\frac{L_{i,j}}{v}
\end{equation}
In this sense, \(v\) represents the expected travel speed of the vehicles. As described in \cite{miletic2022review}, this technique only applies to unidirectional traffic flow optimization. Contrarily, the method \textit{MAXBAND} \cite{little1981maxband} extends the idea of traffic signal coordination by applying mixed-integer linear programming to optimize the offset of adjacent intersections and maximize the bandwidth of the Green Wave in two opposite directions \cite{miletic2022review, yue2021evolution, wang2022critical, wei2019survey, little1981maxband}. In this regard, bandwidth characterizes the duration of the coherent green phase across all signalized locations along a route.

As fixed-time strategies heavily rely on static methods, they are unable to adapt to dynamic traffic changes. In particular, they fail to respond to extraordinary events or unexpected increase in traffic volume, leading to a reduced performance of such a system \cite{miletic2022review,yue2021evolution}. 

\subsection{Actuated Signal Control}
\label{sec:Actuated Signal Control}
As an enhancement to conventional control methods, \textit{Actuated Signal Control} systems can modify the otherwise fixed green phase time, enabling a dynamic response to real-time traffic. Such systems use available road data provided by infrastructure-based sensors (e.g., induction loop detectors or cameras) to identify the current request for green phase extension or termination for any link of an intersection. A well-known algorithm for actuated signal control is \textit{MOVA (Microprocessor Optimised Vehicle Actuation)} \cite{TransportResearchLaboratory.2017}, developed by the Transport Research Laboratory (TRL). Constructed for single and uncoordinated intersections, it operates in two distinct modes: for non-overloaded junctions, it minimizes delays, while for congestion, it optimizes capacity \cite{TransportResearchLaboratory.2017}. Actuated signal control systems are primarily used for isolated intersection control, making Green Wave corridors infeasible due to the enormous variation in cycle times for each intersection, depending on the current traffic situation \cite{miletic2022review,warberg2008green}.

\subsection{Adaptive and Intelligent Signal Control}
\label{sec:Adaptive and Intelligent Signal Control}
To counteract the disadvantages of the aforementioned methods, extensive research work has been conducted in the field of adaptive and intelligent systems. Fundamentally, a distinction must be made between industrial and academic methods, which will be explained in the following.

\subsubsection{Industrial Methods}
The traffic control system \textit{SCOOT (Split Cycle Offset Optimization Technique)} \cite{bretherton1990scoot} was developed by TRL, and is nowadays commonly used in urban areas. As a fully adaptive system, it utilizes data provided by infrastructure sensors to predict changes in traffic flow. As described in \cite{bretherton1990scoot}, \textit{SCOOT} operates on a common cycle time for adjacent intersections, and optimizes the settings for cycle time, green phase durations, and offsets between those.

In addition to this method, \textit{SCATS (Sydney Coordinated Adaptive Traffic System)} \cite{sims1980sydney} represents another adaptive system. According to \cite{sims1980sydney}, it is mainly characterized by its hierarchical control structure, consisting of a supervisor, regional, and local controllers at each intersection. According to this control architecture, the parameters for cycle time, split, and offsets are adjusted in response to the current traffic demand. Like \textit{SCOOT}, this system also uses input from infrastructure sensors to derive an optimal control strategy. 

Also commonly used, \textit{UTOPIA} \cite{mauro1990utopia} represents a hierarchically structured and decentralized control system. As it can be read in \cite{mauro1990utopia}, the system operates on two control levels: the intersection and area levels. At the intersection level, the local controller calculates the settings of the traffic lights at its intersection based on available traffic data while interacting with neighboring controllers. On the other hand, the intersection level communicates with the upper area level controller, predefining constraints for local controllers. 

Since all three methods consider an extended network rather than a single intersection, Green Wave strategies can be effectively implemented with these approaches. In addition, advanced methods like \textit{RHODES} \cite{mirchandani2001real} or \textit{OPAC} \cite{gartner1983opac} have been designed, but show less field penetration rate.

\subsubsection{Academical Methods}
Traffic control strategies based on artificial intelligence, such as \textit{Fuzzy Logic}, \textit{Metaheuristics}, or \textit{Machine Learning}, have gained significant popularity in academia in recent years. Among these, \textit{Fuzzy Logic} has been extensively discussed for traffic control strategies due to its unique advantages, as summarized in \cite{ranpura2022review,agrawal2020intelligent,arifin2021recent}:
\begin{itemize}
    \item Control logic is similar to human logical thinking and thus intuitive and interpretable.
    \item Controllers are flexible and robust, allowing them to handle varying and uncertain traffic conditions.
    \item Control logic uses simple mathematics to model nonlinear system behavior.
\end{itemize}
Although much work has focused on isolated intersections, this method was also extended to the network level. In this sense, Jafari et al. \cite{jafari2022improving} employed the concept of \textit{Fuzzy Logic} to formulate a multi-agent system for coordinating multiple intersections.

A further widely used method is the application of \textit{Metaheuristics}, which mimic natural and biological processes. In the field of signal optimization, these techniques have emerged as inspiring approaches, including evolutionary and swarm intelligence algorithms. Specifically, \textit{Genetic Algorithm (GA)}, \textit{Particle Swarm Optimization (PSO)} and \textit{Ant Colony Optimization (ACO)} are the most commonly used methods \cite{jamal2021metaheuristics,wang2018review,abu2020metaheuristic,qadri2020state}. The following explanations describe these three methods as outlined in \cite{jamal2021metaheuristics,jalili2021application,shaikh2022review}. \textit{Genetic Algorithms} \cite{holland1984genetic} are characterized by an initial population of solution candidates and an objective function to be minimized. These candidates are iteratively combined and mutated according to their so-called fitness until any termination criteria are met. Regarding Green Waves in traffic, Chin et al. \cite{chin2011multiple} demonstrated the successful implementation of such an algorithm for coordinating multiple intersections. Besides, PSO \cite{eberhart1995new} is inspired by the behavior of swarms like birds. A candidate solution of the search space depicts a particle, with the entire population of particles representing a swarm. Each particle successively changes its position and velocity according to its own and the swarm's best solution. Concerning this, Dabiri and Abbas \cite{dabiri2016arterial} applied such a PSO algorithm, improving traffic signal settings of an arterial road with three intersections. Also derived from biological theory, ACO \cite{dorigo2006ant} follows the natural behavior of ants, searching the shortest path from their origin to the food location. Every path is treated as a possible solution to the optimization problem, with better solutions according to the objective to be minimized assigned with higher pheromone concentration. Zechman et al.\cite{zechman2010ant} evaluated the performance of ACO at a one-way and complex network, showing more reliable results than GA.

\begin{figure}[t]
    \centering
    \includegraphics[width=\columnwidth]{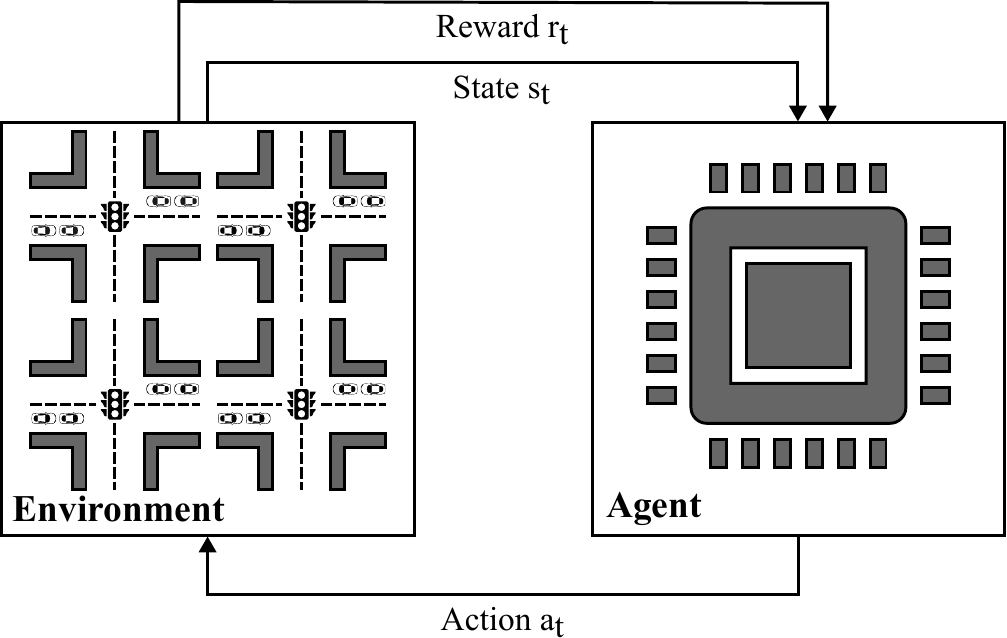}
    \caption{RL Architecture for Traffic Signal Control, based on \cite{haydari2022deep}}
    \label{fig:RL_Architecture}
\end{figure}

An aspiring approach for the design of intelligent traffic lights is the utilization of \textit{Machine Learning}. In particular, \textit{Reinforcement Learning (RL)} has become the subject of much academic work, investigating its potential for single and multiple intersection control. Building  on the explanations in \cite{miletic2022review, haydari2022deep, zhao2024survey, rasheed2020deep, miao2021survey}, the key aspects are highlighted in the following. Using RL, traffic signal control is modeled as a so-called \textit{Markov Decision Process (MDP)}, characterized by the tuple \((S,A,T,R)\) with \cite{puterman1994markov}:
\begin{itemize}
    \item \(S=(s_1,s_2,...,s_n)\) as the set of environment states,
    \item \(A=(a_1,a_2,...,a_m)\) as the set of actions,
    \item \(T(s_{t+1}|s_t,a_t)\) as the transition function, implying the probability of moving from one state \(s_t\) to another \(s_{t+1}\) at time \(t\), and
    \item \(R(s_t,a_t,s_{t+1})\) as the reward function for action \(a_t\) at state \(s_t\) when shifting to state \(s_{t+1}\)
\end{itemize}
The parameters listed in Table \ref{tab:MDP_Parameters} are commonly used for the elements State \(S\), Action \(A\), and Reward \(R\).
\begin{table}[b]
    \centering
    \caption{Parameters of Markov Decision Process \cite{miletic2022review, haydari2022deep, zhao2024survey, rasheed2020deep, miao2021survey}}
    \begin{tabular}{|>{\centering\arraybackslash}m{3.8cm}|
                    >{\centering\arraybackslash}m{3.8cm}|}
    \hline
    \textbf{MDP Element} & \textbf{Possible Parameters}\\
    \hline
    \multirow{4}{*}{\centering State} 
    & Queue length \\ 
    & Average speed \\
    & Average waiting time \\
    & Current traffic signal \\
    \hline
    \multirow{4}{*}{\centering Action} 
    & Phase selection \\ 
    & Phase split \\
    & Phase duration \\
    & Maximum green time \\
    \hline
    \multirow{4}{*}{\centering Reward} 
    & Intersection/Network throughput \\ 
    & Waiting time \\
    & Cumulative delay \\
    & Queue length \\
    \hline
    \end{tabular}
    \label{tab:MDP_Parameters}
\end{table}
The goal of the MDP agent is to find an optimal policy, maximizing the expected cumulative reward. In this respect, Fig. \ref{fig:RL_Architecture} shows the RL architecture with the main components adapted to traffic signal control. If the agent uses or learns an internal representation of the environmental model with its transition probability \(T\), this is called model-based RL. Contrarily, an agent directly tries to find the best policy without any transition model in model-free RL. Basically, there are two main approaches of model-free RL, namely (a) value-based and (b) policy-based RL. With regard to value-based methods, a value function indicates how good a particular state and action is. In this respect, well-known algorithms are Q-learning \cite{watkins1992q} and SARSA \cite{rummery1994line}. On the other hand, policy-based methods like REINFORCE \cite{williams1992simple} directly learn a policy, which iteratively is updated. In high-dimensional spaces, traditional RL reaches its limits. For this reason, \textit{Deep Reinforcement Learning (DLR)} is becoming popular, approximating the optimal policy or value functions by deep neural networks. To manage the coordination of multiple intersections cooperatively, using single agents for individual intersections is less expedient than the application of multi-agent RL. As an example, van der Pol and Oliehoek \cite{van2016coordinated} introduced the combination of Deep Q-learning with a coordination algorithm for multiple-intersection control. Nevertheless, real-world deployment of RL remains challenging as described in \cite{muller2023bridging,chen2022real}.

Although there are additional approaches, such as \textit{Model Predictive Control} \cite{ye2019survey}, this work focuses primarily on the methods discussed here, as they characterize the main findings of the selected papers.

\section{Trends, Potential of Green Wave Technology and its Impact on V2X, Environment, and Safety}
\label{sec:challenges}

The incorporation of green wave technology into traffic management systems has significantly improved traffic flow. To ensure the section's focus and relevance, this section briefly discusses the works that directly address the topic of the green wave and its impact on V2X communication, environmental sustainability, and safety. 


\subsection{V2X Integration and Technical Advancements}
\label{sec:V2X Integration and Technical Advancements}

Cellular-V2X (C-V2X) integration for effective communication has completely reshaped the Green Wave implementations in the real world \cite{chochliouros2021c}. Communication between vehicles and traffic signal systems can be achieved wirelessly using technologies like IEEE 802.11p or 5G C-V2X \cite{chochliouros2021c}. These transformative developments facilitate the building of an interconnected environment through consistent communication among all transportation network elements, enabling robust decision-making and flexible network traffic management. The effective use of advanced techniques like the Multi-Intersection Coordination Algorithm with V2X (MICA-V) and C-V2X in multiple intersection coordination is highlighted by the authors in \cite{partani2021achievable,shi2020coordination,chen2016cooperative}. These systems utilize onboard units (OBUs), roadside units (RSUs), and a central coordination module to ensure the synchronization of traditional traffic lights across all junctions, achieving a smooth traffic flow without interruptions or stops. Nevertheless, the traditional solution fails to cope with shockwaves and congestion due to the growing complexity of dynamic traffic. Such new traffic dynamics depend significantly on the efficient sharing of real-time data between vehicles and infrastructure. V2X, being the backbone, ensures that the Green Wave solutions can facilitate higher traffic volumes and minimize travel delays. 

To overcome the earlier limitations, \cite{zheng2020novel} and \cite{ma2019green} have introduced an adaptive traffic signal control strategy. This solution combines reinforcement learning and V2X communications to enable valuable real-time decision-making. Moreover, dynamic traffic management is achieved by integrating a multi-agent environment with edge computing for precise real-time signal timing adjustments based on traffic conditions \cite{wang2021collaborative}.  The use of Deep Q-Network (DQN), a value-based reinforcement learning algorithm, in these applications not only improves signal coordination but also enhances vehicle flow \cite{li2024green}.  Real-time data exchange through V2X improves the adaptability and learning efficiency of these models, making the solutions robust for traffic management. 

Furthermore, C-V2X enables proactive platooning for seamless, uninterrupted, and optimal traffic flow at junctions with traffic signals through networked control and computation-centralized platooning \cite{gao2023proactive}. Urban congestion is significantly reduced through platooning, where a group of vehicles coordinates by maintaining suggested speeds for passing through the Green Waves without unnecessary stopping \cite{wang2023research}, \cite{yang2020study}. Sophisticated algorithms rely on factors such as driver behavior, vehicle type, capabilities, uncertain incidents, and road conditions to form and maintain smooth platooning. Past findings have shown that traffic flow efficiency improves due to vehicle coordination for platooning, which, in turn, supports current smart traffic management systems more efficiently. V2X also facilitates the inclusion of eco-driving schemes \cite{kui2019eco,yang2020study} to reduce the vehicle's fuel consumption and carbon emissions. By doing so, Platooning and eco-driving enhance the potential of Green Wave systems.

As a further supportive technology, Greenlight Optimal Speed Advisory (GLOSA) is widely used for effective traffic and speed management \cite{suzuki2020green}. It uses real-time data from traffic lights, vehicles, traffic intensity, weather information, and infrastructure to guide drivers with appropriate speeds for seamless driving through green signal corridors and optimize vehicle trajectories \cite{yang2020study,chochliouros2021c,islam2018impact}. Fuel savings of up to 15\% are achieved through advanced GLOSA techniques \cite{de2019green}, making it suitable for improving environmental conditions.

DDPG-BAND combines the Deep Deterministic Policy Gradient (DDPG) algorithm with a bandwidth-oriented reward function to optimize signal timing across intersections \cite{zheng2020novel}. ES-BAND leverages an evolutionary strategy (ES) to iteratively improve signal timing plans by simulating traffic flow and selecting configurations that maximize Green Wave bandwidth \cite{zheng2020novel}. Adaptive algorithms, such as DDPG-BAND and ES-BAND, support dynamic Green Wave through V2X. The algorithm’s adaptability and effective scalability ensure exceptional management of complex traffic networks for massive urban infrastructures. Integrating reinforcement learning, eco-driving, multi-agent systems, and edge computing with V2X will ensure the future of smart cities with scalable, sustainable, and efficient transportation systems. The practical benefits of V2X in Green Wave are summarized in detail in Table \ref{tab:v2x_analysis}, which includes notable improvements in travel time, energy efficiency, and overall traffic flow.

\begin{table*}[ht]
\renewcommand{\arraystretch}{1.2} 
\setlength{\tabcolsep}{3pt} 
\centering
\caption{Potential of V2X-based Green Wave Implementation and Supporting Technologies}
\label{tab:v2x_analysis}
\begin{tabular}{|c|c|p{2.2cm}|p{3.2cm}|p{4.0cm}|p{3.5cm}|p{2.0cm}|}
\hline
\textbf{[Ref]} & \textbf{Year} & \textbf{Primary Focus} & \textbf{Technologies (Tech.) Used} & \textbf{Key Implementation Features} & \textbf{Outcomes} & \textbf{Application } \\ \hline
\cite{chen2016cooperative} & 2016 & Green Wave Coordination & \textbf{V2X Concept}: V2V, V2R \newline
\textbf{Communication Tech.}: IEEE 802.11p (WAVE / DSRC) & - Cooperative throughput maximization \newline - Green Wave coordination algorithms \newline - Traffic control across multiple intersections & - Increased global throughput \newline - Reduced waiting and travel time \newline - Lower CO2 emissions & Arterial roads \newline with multiple intersections \\ \hline
\cite{li2024green} & 2024 & Ecological Speed Planning & \textbf{V2X Concept}: V2N, I2N \newline
\textbf{Communication Tech.}: 4G, 5G, DSRC & - Two-stage optimization strategy \newline - DQN for queue estimation \newline - Multi-intersection objective \& constraints strategy & - 16.65\% energy reduction \newline - 26.33\% travel time improvement  & Multi-intersections \newline with EVs \\ \hline
\cite{kui2019eco} & 2019 & Eco-speed Optimization & \textbf{V2X Concept}: V2I \newline
\textbf{Communication Tech.}: Not specified & - Eco-speed optimization \newline - Fuel consumption reduction \newline - Multi-intersection speed optimization & - 20\% fuel reduction \newline - Improved efficiency \newline - Smoother flow & Signalized intersections \\ \hline
\cite{gao2023proactive} & 2023 & Proactive Platooning & \textbf{V2X Concept}: V2V, V2I \newline
\textbf{Communication Tech.}: C-V2X (PC5 interface) & - Dynamic platoons \newline - Real-time coordination \newline - Adaptive control \newline - Enhanced stability & - Delays reduced \newline - Improved throughput \newline - Platoon stability +45\% & Urban intersections \\ \hline
\cite{chochliouros2021c} & 2021 & GLOSA for Green Wave & \textbf{V2X Concept}: V2N, V2V, V2I \newline
\textbf{Communication Tech.}: 5G & - GLOSA for speed optimization \newline - Real-time traffic light information \newline - Vehicles advised on optimal speeds & - Improved traffic flow \newline - Reduced fuel consumption \newline - Enhanced driving comfort & Intersections with \newline GLOSA-based speed optimization \\ \hline
\end{tabular}
\vspace{0.5em}
\begin{flushleft}
\textit{* Abbrevations:} V2N – Vehicle-to-Network, I2N – Infrastructure-to-Network, PC5 – Direct sidelink interface for C-V2X communication, WAVE – Wireless Access in Vehicular Environments, DSRC – Dedicated Short-Range Communications.
\end{flushleft}
\end{table*}

\subsection{Environmental Impact and Sustainability}
\label{sec:V2X Environmental Impact and Sustainability}
\begin{figure}[b]
    \centering
    \includegraphics[width=0.9\linewidth]{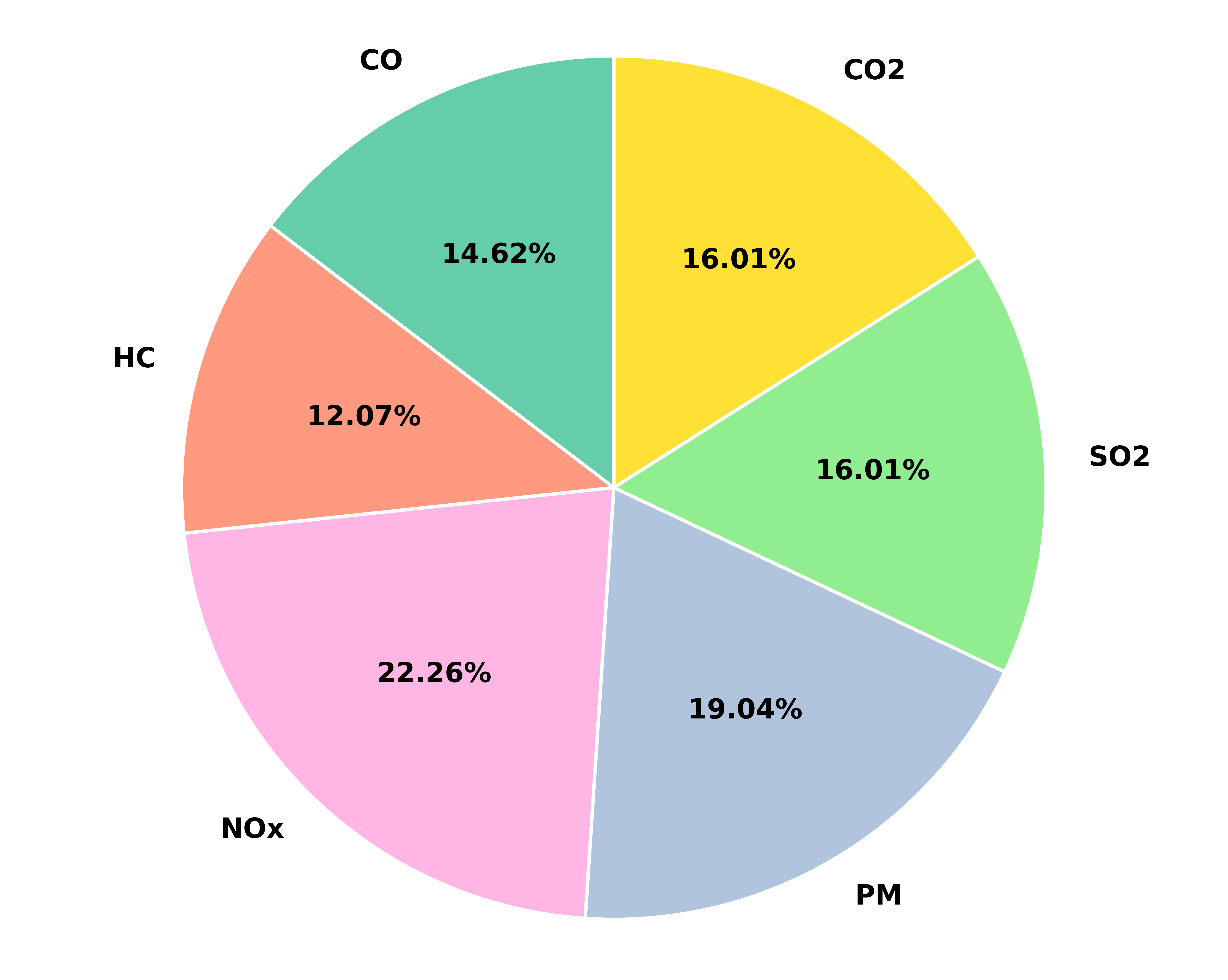} 
    \caption{Percentage Reduction in Emissions with Green Wave \cite{kiers2017effect}.}
    \label{fig:percentage_reduction_emissions} 
\end{figure}

Reducing vehicle emissions has gained significant momentum by utilizing speed-coordinated algorithms integrated with V2X technologies \cite{paul2020influence}. 
However, signalized intersections often increase CO2 emissions by 11.52\% due to red light stops, with Carbon Monoxide (CO), Hydrocarbons (HC), and Nitrogen Oxides (NOx) emissions rising by 16.46\%, 16.89\%, and 15.72\%, respectively, emphasizing the need for better signal timing \cite{zhao2019estimation}. Optimized signal timing in mixed traffic, including fuel and electric vehicles, effectively reduces emissions during acceleration, deceleration, and idling stages \cite{zhao2021modeling, fazzini2022effects}. Adaptive traffic light control systems using reinforcement learning significantly enhance traffic flow, achieving a 41\% increase in average velocity while reducing CO2 emissions, promoting sustainability in urban mobility \cite{koch2023adaptive}. Studies show that the well-synchronized routes on urban arterial roads, Green Waves, can reduce emissions like CO2, NOx, and PM10 by 10\% to 40\% under optimal conditions, significantly decreasing the environmental impact of traffic \cite{de2011traffic}. In another study \cite{kiers2017effect}, Green Waves lower CO2 and other pollutant emissions, as shown in Figure \ref{fig:percentage_reduction_emissions}, countering stop-and-go traffic.

The electric vehicle (EV) infrastructure, incorporating the Green Wave technology, led to new possibilities for optimizing the environment \cite{zhao2021modeling}. These EVs can achieve maximum energy efficiency by proper coordination between traffic lights, dynamic route planning systems, and smart charging stations. Moreover, for EVs to optimize vehicle routing and charging routines, these technology integrations consider various factors, such as battery charging level, availability of charging stations, and traffic conditions. Such an integrated system reduces the environmental impact, thereby enhancing more sustainable urban mobility solutions \cite{jochem2015assessing}.

\subsection{Safety Considerations and Risk Management}
\label{sec:Safety Considerations and Risk Management}
Advanced V2X systems optimize traffic flow and reduce delay through Green Wave solutions, but considering the safety of all road users becomes crucial for building efficient and successful systems. Algorithms like DQN are used in emergency vehicle prioritization \cite{cao2022gain} and Green Wave guidance for human drivers \cite{yuan2022deep} to ensure urgent traffic is prioritized and vehicles are smoothly passed through intersections. Although they effectively reduce congestion and delays, \cite{biswas2017effect} highlights the issue of having countdown timers where red-light violations may affect vehicle-pedestrian coordination. Similar to Green Waves, synchronization through adaptive signal control using image-sensing software is crucial for improving user safety and avoiding accidents due to information delays and inaccurate data processing \cite{cheewapattananuwong2011mitigating}. 

Furthermore, effective risk management is crucial for achieving seamless traffic flow and prioritizing emergency vehicles. \cite{wang2021green} and \cite{bieker2019modelling} utilize big data analysis to optimize signal timings; however, precise data integration is crucial to avoid errors by pedestrians, cyclists, and drivers \cite{fickas2021green, de2019green}. Studies on Green Signal Countdown Timers (GSCTs) show their capability to reduce red-light running \cite{pan2023impact, paul2020influence, brand2024riding}, while situations like aggressive driver behaviors and unpredictable driver response require sophisticated risk management strategies \cite{islam2017safer}.

\section{Implementation Challenges and future directions}
The latest developments in coordinated traffic control strategies reveal a clear shift to the RL algorithm to handle dynamic and complex traffic situations. However, all approaches associated with RL require high computational resources. V2X has paved the way for numerous advancements in Green Wave systems; however, challenges remain regarding real-time usage and deployment. With the increasing complexity of traffic networks, real-time coordination is becoming more challenging, leading to scalability issues. Issues such as data packet loss, delays, network bottlenecks, and the penetration rate of V2X technology can affect the system's performance. Additionally, uncertain driver behavior and dynamic traffic scenarios can complicate the optimization of traffic flow. Moreover, ensuring safety through the design of such solutions for vulnerable road users, such as pedestrians, is crucial and complicated due to broader system expansions. 

Future research will focus on addressing these challenges and limitations through advanced data processing, edge computing, 5G networks, and autonomous vehicles, which will further enhance the adaptation of the Green Wave system. Inter-modal traffic management and advanced privacy, cybersecurity, and safety protocols for all road users are crucial for building a safer and more robust traffic management system integrated with Green Wave technologies. Furthermore, another significant challenge of Green Wave is finding common strategies for different traffic participants, including both motorized and vulnerable road users, such as vehicles, pedestrians, and cyclists, due to their varying speeds, acceleration profiles, and lane usage patterns. Achieving a trade-off between fairness and efficiency in signal coordination requires advanced traffic modeling, where priority rules or adaptive speed recommendations are tailored to each vehicle type and road user group.

\section{Conclusion}
\label{sec:conclusion}
This work presents a comprehensive survey on the challenges and advantages of using Green Wave technology for optimal urban traffic management. Although significant advancements have been made in improving traffic flow, reducing congestion, and decreasing emissions through the integration of Green Wave with advanced technologies like V2X and edge computing, deploying in real-time and coordinating across multiple intersections remains challenging for effectively handling dynamic traffic conditions. The importance of safety measures and risk mitigation for vulnerable road users is also highlighted in this paper through a briefing about both opportunities and risks involved in Green Wave. 

Adaptive, intelligent signal control strategies and eco-driving models have been identified as promising solutions; however, a notable research gap still exists regarding real-time scalability and proactive decision-making in dense urban environments. This survey emphasizes the need for future research to focus on Green Wave, including cybersecurity for information protection and ensuring seamless communication through the latest V2X, edge computing, and 5G-based solutions, before integrating them into smart city infrastructures.  These solutions would transform urban mobility, making transport management systems more robust, sustainable, and efficient for every road user. In general, the authors propose to evaluate the effectiveness of Green Wave technology using available real-world traffic datasets, such as the one provided for the city of Darmstadt \cite{VerkehrsdatenDarmstadt}.

\bibliographystyle{IEEEtran}
\bibliography{references}

\end{document}